\documentclass[aps,prl,nofootinbib,amsmath,amssymb,twocolumn,10pt]{revtex4}

\pdfoutput=1

\usepackage{graphicx}
\usepackage{dcolumn}
\usepackage{bm}
\usepackage{amssymb}
\usepackage{latexsym}
\usepackage{booktabs}
\usepackage{amsmath}
\usepackage{multirow}
\usepackage[colorlinks=true, linkcolor=red, citecolor=blue]{hyperref}

\newcommand{\be}{\begin{equation}}
\newcommand{\ee}{\end{equation}}
\newcommand{\bq}{\begin{eqnarray}}
\newcommand{\eq}{\end{eqnarray}}

\begin{document}

\title{Impact of the latest measurement of Hubble constant on constraining inflation models}

\author{Xin Zhang}
\email{zhangxin@mail.neu.edu.cn}
\affiliation{Department of Physics, College of Sciences,
Northeastern University, Shenyang 110004, China}
\affiliation{Center for High Energy Physics, Peking University, Beijing 100080, China}

\begin{abstract}
We investigate how the constraint results of inflation models are affected by considering the latest local measurement of $H_0$ in the global fit. We use the observational data, including the Planck CMB full data, the BICEP2 and Keck Array CMB B-mode data, the BAO data, and the latest measurement of Hubble constant, to constrain the $\Lambda$CDM+$r$+$N_{\rm eff}$ model, and the obtained 1$\sigma$ and 2$\sigma$ contours of $(n_s, r)$ are compared to the theoretical predictions of selected inflationary models. We find that, in this fit, the scale invariance is only excluded at the 3.3$\sigma$ level, and $\Delta N_{\rm eff}>0$ is favored at the 1.6$\sigma$ level. The natural inflation model is now excluded at more than 2$\sigma$ level; the Starobinsky $R^2$ model becomes only favored at around 2$\sigma$ level; the most favored model becomes the spontaneously broken SUSY inflation model; and, the brane inflation model is also well consistent with the current data, in this case.

\end{abstract}

 \maketitle

Recently, Riess et al.~\cite{Riess:2016jrr} reported the new result of local measurement of the Hubble constant, $H_0=73.00\pm 1.75$ km s$^{-1}$ Mpc$^{-1}$, which is 3.3$\sigma$ higher than the fit result of $66.93\pm 0.62$ km s$^{-1}$ Mpc$^{-1}$ derived by the Planck collaboration based on the $\Lambda$CDM model with $\sum m_\nu=0.06$ eV using the latest Planck CMB data. The tension between the latest $H_0$ measurement and the Planck data has inspired numerous discussions. On one hand, it might be caused by some systematic uncertainties in the measurements. On the other hand, perhaps one has omitted some unknown physical factors in the cosmological model, which leads to some inconsistencies among different data sets. For example, replacing the cosmological constant with a dynamical dark energy \cite{de} and considering the extra relativistic degrees of freedom (i.e., an additional parameter $N_{\rm eff}$) \cite{Riess:2016jrr,Zhang:2014dxk,Zhang:2014ifa} both can help relieve this tension to some extent. It was also shown in Ref.~\cite{Zhang:2014ifa} that the involvement of light sterile neutrinos in the cosmological model can simultaneously relieve almost all the tensions among the current astrophysical observations, which leads to a new cosmic concordance. 

The new measurement of $H_0$ can play an significant role in the cosmological global fit because the uncertainty of this result has already been reduced from 3.3\% to 2.4\% by using the Wide Field Camera 3 (WFC3) on the Hubble Space Telescope (HST) \cite{Riess:2016jrr}. But, once the latest $H_0$ measurement is combined with other astronomical observations, one must be aware of the importance of trying to make these data consistent in the analysis. 

The Planck collaboration \cite{Ade:2015xua} released their latest data of the CMB anisotropies on both temperature and polarization based on the full Planck survey, which measure the spectral index of curvature perturbations to be $n_s=0.968\pm0.006$ and its scale dependence to be $dn_s/d\ln k=-0.003\pm 0.007$, when combined with the Planck lensing likelihood. The upper bound on the tensor-to-scalar ratio is derived to be $r_{0.002}<0.11$ (95\% CL). These measurements tightly constrain the inflation models (see Fig.~12 of Ref.~\cite{Ade:2015lrj}). Also, recently, the Keck Array and BICEP2 collaborations \cite{Array:2015xqh} released the first Keck Array CMB B-mode polarization data at 95 GHz, and they presented the results from an analysis of all data taken by the BICEP2 and Keck Array CMB polarization experiments up to and including the 2014 observing season. Using the B-mode data of BICEP2 and Keck Array yields an upper bound, $r_{0.05}<0.09$ (95\% CL). Once the BICEP2/Keck B-mode data are combined with the Planck CMB data plus other astrophysical observations, a much tighter limit of $r$ is obtained, $r_{0.05}<0.07$ (95\% CL). Adding the BICEP2/Keck data to the Planck cosmological global fit further improves the constraints on the inflation models (see Fig.~7 of Ref.~\cite{Array:2015xqh}). 

Furthermore, in Ref.~\cite{Huang:2015cke}, by considering the latest baryon acoustic oscillation (BAO) data from the Data Release 12 (DR12) of SDSS-III BOSS in the cosmological fit, the authors showed that the constraints on inflation are further improved (see Fig.~4 of Ref.~\cite{Huang:2015cke}). They show that the inflation models with a convex potential is not favored and both the natural inflation model and the models with a monomial potential are only marginally favored at around 95\% CL; but the Starobinsky inflation model can fit these data quite well.

\begin{figure}[!htbp]
  \includegraphics[width=8.5cm]{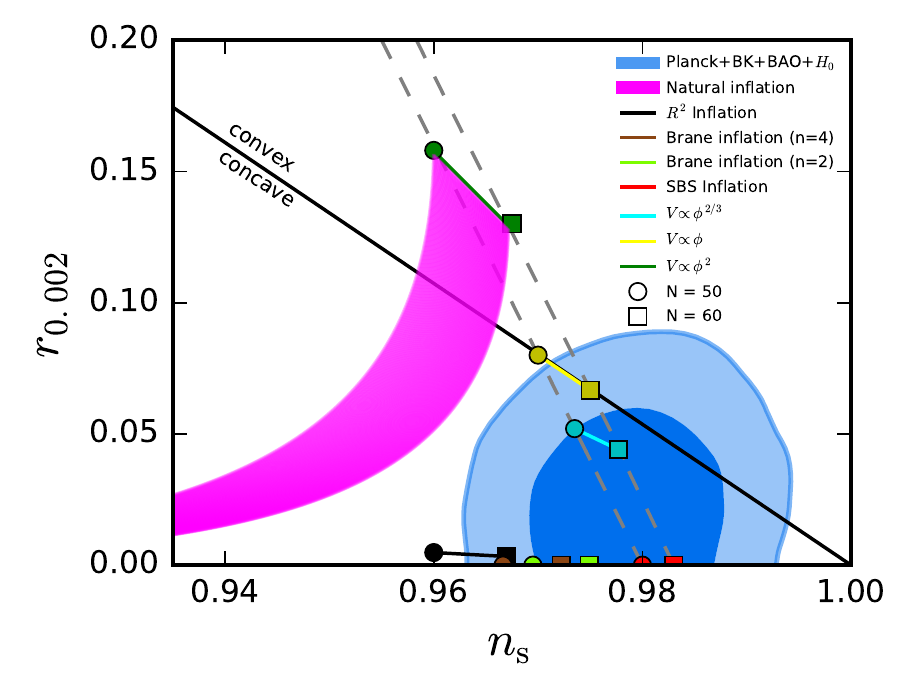}
  \caption{The marginalized joint 68\% and 95\% CL contours of $(n_s, r)$ at $k=0.002$ Mpc$^{-1}$ in the $\Lambda$CDM+$r$+$N_{\rm eff}$ model using the Planck+BK+BAO+$H_0$ data, compared to the theoretical predictions of selected inflationary models.}\label{fig1}
\end{figure}

Here we will investigate how the constraint results of inflation models are affected by further considering the latest local measurement of $H_0$ in the global fit. The observational data used in this work include: the Planck CMB full data (Planck TT, TE, EE + lowP + lensing, denoted as ``Planck'') \cite{Ade:2015xua}, the BICEP2 and Keck Array CMB B-mode data (denoted as ``BK'') \cite{Array:2015xqh}, the BAO data including the measurements from 6dFGS, MGS, and BOSS DR12 (CMASS and LOWZ) (denoted as ``BAO'') \cite{bao}, and the latest measurement of Hubble constant (denoted as ``$H_0$'') \cite{Riess:2016jrr}. It is known that the BK and BAO data are consistent with the Planck data in the $\Lambda$CDM+$r$ model. But the $H_0$ measurement is in tension with these data, which leads to the problem of how to involve the $H_0$ measurement in the fit. Since introducing the extra relativistic degrees of freedom (additional parameter $N_{\rm eff}$) in the cosmological model can effectively relieve the tension, as mentioned above, we decide to consider the $\Lambda$CDM+$r$+$N_{\rm eff}$ model in this work. Thus, now, the observational data in the analysis are basically consistent. 

The marginalized joint 68\% and 95\% CL contours of $(n_s, r)$ at $k=0.002$ Mpc$^{-1}$ in the $\Lambda$CDM+$r$+$N_{\rm eff}$ model using the Planck+BK+BAO+$H_0$ data are shown in Fig.~\ref{fig1}. Here we have assumed $dn_s/d\ln k=0$. And, the consistency relation, $r=-8n_t$, which will be held when inflation is driven by a single slowly-rolling scalar field with a standard kinetic term, is used in the analysis. The theoretical predictions of selected inflationary models are compared to the observational constraints of $(n_s, r)$ in Fig.~\ref{fig1}. The slow-roll approximation is applied to these models. 

In this fit, we obtain $n_s=0.9787^{+0.0064}_{-0.0065}$, only excluding the Harrison-Zel'dovich scale-invariant spectrum at the 3.3$\sigma$ level, and we obtain the upper bound $r_{0.002}<0.071$ (95\% CL). The combined data also give the fit result of the extra relativistic degrees of freedom (also called dark radiation), $N_{\rm eff}=3.30\pm0.16$, which indicates that $\Delta N_{\rm eff}>0$ (here, $\Delta N_{\rm eff}\equiv N_{\rm eff}-3.046$) is favored at the 1.6$\sigma$ level. The minimal $\chi^2$ of this fit is $\chi^2_{\rm min}=13612.184$.

In Fig.~\ref{fig1}, we compare the predictions of the selected inflationary models with the fit results of $(n_s, r)$ in this analysis. We show that, in this case, both convex and concave potentials are favored at 2$\sigma$ level, though the concave potential is more favored. The natural inflation model is now excluded by the data at more than 2$\sigma$ level. For the inflation models with a monomial potential, we show that the $\phi^2$ model is totally excluded, the $\phi$ model is only marginally favored (in the 2$\sigma$ region), and the $\phi^{2/3}$ model is well favored (in the 1$\sigma$ region). In the current situation, the Starobinsky $R^2$ model becomes only favored at around 2$\sigma$ level (on the edge of the 2$\sigma$ region, and the $N=50$ point even lies out of the region), while the most favored model becomes the spontaneously broken SUSY (SBS) inflation model, which locates at the center of the contours. In addition, the brane inflation model (with both $n=2$ and $n=4$; see Ref.~\cite{Ma:2013xma} for comparison) is also well consistent with the current data in this case. 

In this Letter, we give a brief analysis on the impact of the latest measurement of $H_0$ on the inflation model constraints. The comprehensive, detailed analysis on this issue will be left to a forthcoming longer paper.

\begin{acknowledgments}
This work was supported by the National Natural Science Foundation of China (Grants No.~11522540 and No.~11690021), the Top-Notch Young Talents Program of China, and the Provincial Department of Education of Liaoning (Grant No.~L2012087).

\end{acknowledgments}

\end{document}